\documentclass[manuscript]{aastex}

\begin{document}
\title{Chandra detection of X-ray Absorption Associated with a Damped
Ly$\alpha$ System} 
\author{Jill Bechtold\altaffilmark{1}, Aneta Siemiginowska\altaffilmark{2},
Thomas L. Aldcroft\altaffilmark{2}, Martin Elvis\altaffilmark{2}, 
Adam Dobrzycki\altaffilmark{2}
}

\altaffiltext{1}{Steward Observatory, University of Arizona, 933 N. Cherry 
Avenue, Tucson AZ 85721;
jbechtold@as.arizona.edu}
\altaffiltext{2}{High Energy Division, Harvard-Smithsonian Center for Astrophysics; aneta,aldcroft,elvis,adam@cfa.harvard.edu}

\received{}
\accepted{}

\begin{abstract}
We have observed three quasars, PKS 1127-145, Q 1331+171 and Q
0054+144, with the ACIS-S aboard the Chandra X-ray Observatory, in
order to measure soft X-ray absorption associated with intervening
21-cm and damped Ly$\alpha$ absorbers. For PKS 1127-145, we detect
absorption which, 
if associated with an intervening $z_{abs}$=0.312 absorber,
implies a metallicity of 23$\%$ solar. If the absorption is not
at $z_{abs}$=0.312, then the metallicity is still constrained to be 
less than 23$\%$ solar.
The advantage of the X-ray
measurement is that the derived metallicity is insensitive to
ionization, inclusion of an atom in a molecule, or depletion onto
grains.  The X-ray absorption is mostly due to oxygen, and is consistent with
the oxygen abundance of 30$\%$ solar derived from optical nebular emission
lines in a foreground galaxy at the redshift of the absorber.  For Q
1331+171 and Q 0054+144, only upper limits were obtained, although the
exposure times were intentionally short, since for these two objects
we were interested primarily in measuring flux levels to plan for
future observations.  The imaging results are presented in a companion
paper.

\end{abstract}


\section{Introduction}

The optical-ultraviolet spectra of quasars show narrow absorption 
lines from a range of heavy elements which originate in the gaseous
disks or halos of intervening galaxies (reviewed by 
Bechtold 2001).  The highest column absorbers -- the damped Ly-$\alpha$ 
systems and 21-cm absorbers -- are thought to originate in the 
progenitors of large galaxies like the Milky Way at high redshift 
(Wolfe, Lanzetta, Foltz, \& Chaffee 1995; Rao, Turnshek, \& Briggs 1995;
Storrie-Lombardi \& Wolfe 2000 and references therein).  
At moderate redshift (z$<$1) they
generally appear associated with spirals or low surface brightness
galaxies (Steidel, Dickinson $\&$ Persson 1994; Steidel, Pettini, 
Dickinson $\&$ Persson 1994; Steidel et al. 1997; Le Brun, Bergeron, Boisse, \& 
Deharveng 1997; Lane et al. 1998; Boisse et al. 1998).
The damped Ly$\alpha$ absorbers have been the subject of many 
detailed studies of the metallicity of the absorbing gas, 
since the damping profile of Ly-$\alpha$ 
yields an accurate measure of the H I column density, a 
first step in measuring abundances, and the relatively high columns 
enable weak, unsaturated transitions of important ions to be detected.
For some damped Ly$\alpha$ systems, 21-cm absorption can be detected
as well; individual 21-cm velocity components 
can often be deblended more easily than the UV absorption components, 
allowing kinematics of the parent galaxy to be probed (Wolfe \& Briggs 1981; Briggs \& Wolfe 1983; Carilli et al. 1996; Lane, Briggs, \& Smette 2000) 

The detailed abundance pattern of the heavy elements in quasar 
absorbers reflect the nucleosynthetic history of the parent galaxy, 
and hence the origin of the gas.  The interpretation of the UV 
absorption spectrum,
however, is uncertain,  despite very high quality spectra in many
cases (Lu et al. 1996; Prochaska \& Wolfe 1999; Pettini et al. 2000
and references therein).  
The problem is disentangling the effect of dust depletion from 
intrinsic metallicity variations, in the face of large uncertainties due to
saturation of the lines of the most abundant elements, and the uncertain
ionization corrections for the abundant elements.

In this paper, we report the first results of a program to measure
abundances in quasar absorbers in a new way, namely by measuring
the photoelectric absorption in the soft X-rays.  The advantage
of the X-ray absorption is that it will give a measure of the 
heavy element column density that is essentially independent of
ionization, inclusion of an atom in a molecule, or depletion onto
grains (Morrison \& McCammon 1983; Wilms, Allen \& McCray 2000).  
Here, we describe Chandra
ACIS-S observations of two 21-cm and damped Ly$\alpha$ absorbers,
at $z=0.312$ in PKS 1127-145 ($z_{em}$=1.17), $z=1.77$ in 
Q 1331+171 ($z_{em}$=2.08), and a third candidate damped Ly$\alpha$ 
absorber, Q 0054+1331 ($z_{em}$=0.171).  
The imaging of PKS 1127-145 is presented 
in a companion paper (Siemiginowska et al. 2001).

\section{Observations and Analysis}

We observed the quasars with a back-side illuminated CCD, ACIS-S (CCD
number 7) on board the Chandra X-ray Observatory 
(Weisskopf et al. 1996; 2000).  Details of
the observations are listed in Table 1.  We read out custom subarrays
(1/8 subarray) and moved the target off-axis in order to mitigate
$``$pile-up", which distorts the X-ray spectrum.  
The measured count rates indicate that pile-up is
negligible: less than 1$\%$ of the source photons are
affected (McNamara, 2001).  For all
objects, the position of the X-ray source agreed with the optical
positions to better than 1 $\arcsec$.  The right ascension and
declination of the X-ray sources are listed in Table 1.

The data were reduced by the standard pipeline (version R4CU5UPD7.1
with CALDB v.1.7 for Q~0054+1331 and version R4CU5UPD13.3 with CALDB
2.1 for PKS~1127-145 and Q~1331+171). CALDB~2.1 was used to create the
instrument responses and PSF files. We used CIAO 2.0 and Sherpa to
analyze the data, and cross-checked the fits with XSPEC where feasible.

Source and background spectra were extracted from the event files
assuming circular or elliptical regions for the quasars with {\tt
dmextract} and then fit in Sherpa.  (Data for Q1331+171 were obtained
far off-axis so an elliptical source region was required).  

Results of spectral fitting are
given in Table 2.  We fit each spectrum with a power law, with fixed
unredshifted absorption with the Galactic column, and redshifted absorbers,
as indicated in Table 2.  We did not use data with E$<$ 0.4 keV since large
uncertainties exist in the calibration for soft energies at this time.
We discuss each object in turn.

\section{Q 0054+144, PHL 909, $z_{em}=0.171$}

Q~0054+144 is a low redshift QSO, which, prior to the launch of
Chandra, was known from $ROSAT$ 
observations to be bright in the X-rays (HEASARC ROSAT archive,
Snowdon et al. 1995).  It is radio quiet, and
has a well-studied host galaxy, which has colors and morphology
suggesting that it is an early-type galaxy (Bahcall, Kirhakos \& 
Schneider 1996; McLure et al. 1999).  A candidate damped
Ly$\alpha$ absorber at $z_{abs}=0.102$ was suggested by Lanzetta
et al. (1995) using $IUE$ data.  
Because Q~0054+144 appeared to be bright enough to
allow a high quality grating observation with $Chandra$, 
we obtained a short exposure
to measure the spectral energy distribution precisely, particularly in
the critical soft X-ray region where the oxygen edges are expected.
The results of the $Chandra$ observation spectral fits are listed in
Table 2, and the best fit power law is shown in Figure 1. 

The $Chandra$ 
X-ray spectrum is adequately fit with a power law and no intervening
absorption. 
Subsequently, we became aware of a Hubble Space Telescope GHRS spectrum of Q 0054+144 taken
with the G140L grating, which showed that no damped Ly$\alpha$
absorption is present at $z_{abs}=0.102$.  The GHRS
spectrum is shown in Figure 2.  The most prominent absorption feature is a
complex of absorption at the emission line redshift, seen in
Ly$\alpha$ and N V $\lambda\lambda$1238,1242.  Si II and Si III are
weak or absent, suggesting that the system originates with material
intrinsic to the quasar, although O VI is also absent.
A low ionization metal absorption line system is
present at $z=0.102$, but Ly$\alpha$ is not damped. 
Lines at 1354 and 1365~$\AA$ are 
probably Ly$\alpha$ lines at $z=0.114$ and 0.123,
with metal absorption weak or absent.

Despite the lack of an intervening damped Ly$\alpha$ absorber or 
a strong X-ray absorber in the Chandra spectrum, Q0054+144 
would be a good candidate for future X-ray grating 
observations, because of its high count rate, steep X-ray spectrum and
associated ($z_{abs}\sim z_{em}$) ultraviolet absorption system.

\section{Q1331+170, $z_{em}=2.08$}

Prior to our observations, Q 1331+170 had never been observed before
in the X-rays, even though optically it is one of the brightest $z=2$
quasars known.  There is a strong damped Ly$\alpha$ and 21-cm absorber at
$z=1.77$ which contains unusually low ionization material, including C
I (Carswell et al. 1979; Wolfe \& Davis 1979; Songaila et al. 1994).

The short (3000 second) 
$Chandra$ observation yielded only 198 counts, and a 3-$\sigma$
upper limit to N$_H$~  at $z_{abs}$=1.77 of 1.9$\times$ 10$^{22}$
cm$^{-2}$, assuming solar abundance. This value is larger than the
observed N(HI) from Ly$\alpha$ of 1.6$\times$10$^{21}$ cm$^{-2}$.
With a longer exposure, X-ray absorption should be detectable.

\section{PKS 1127-145, $z_{em}=1.18$}

PKS 1127-145 is a bright, gigahertz-peaked radio source with a VLBI
and X-ray jet (Kellerman et al. 1998; Siemiginowska et al. 2001).  The
properties of the X-ray jet are discussed in the companion paper
(Siemiginowska et al. 2001).  A strong intervening Mg II and Mg I
absorption system at $z_{abs}=0.312$ was discovered by Bergeron \& Boisse
(1991) who also identified two late type galaxies near the quasar line
of sight at the redshift of the absorber.  The nearer galaxy,
separated by 9.6 $\arcsec$ from the quasar (27 h$^{-1}_{100}$ kpc),
has strong [O II], [O III] and H$\beta$ emission lines.  An
observation with FOS of the galaxy by Deharveng et al. (1995) did not
detect Ly$\alpha$ emission, and although this galaxy is one of the
reddest in the survey of Steidel (1993) it appears to have little
extinction.  Deharveng et al. (1995) then estimated that the
oxygen abundance is approximately 30$\%$ solar, based on the
nebular emission line ratios.

Lane et al. (1998) discovered H I 21-cm absorption at $z=0.3127$ 
and described the Ly$\alpha$ profile of the absorber, based on FOS
data, which appears damped.  The neutral hydrogen column density
derived from the damped Ly$\alpha$ profile is N(HI)=(5.1$\pm$0.9)$\times$
10$^{21}$ cm$^{-2}$.  Further, they obtained an optical spectrum of
a third galaxy with [O III] emission lines, and a 
separation, 3.9 $\arcsec$, or 11 h$^{-1}_{100}$ kpc from the quasar.
This may be the absorber, or the absorption could arise in tidal
debris associated with this group of galaxies.

The X-ray spectrum of PKS 1127-145 shows significant absorption 
in excess of the Galactic value.  Figure 3a shows the data with
E$>$2 keV fit 
by a power law and Galactic absorption; the fit extrapolated for 
E$<$2 keV is poor, 
with systematic residuals.  Figure 3b shows the data fit by
a power law, Galactic absorption, and intervening absorption
at the redshift of the 21-cm absorber, $z=0.312$.   

The statistics are good enough to show significant
features in the residuals.  All of the features can be accounted for
by calibration uncertainties, that is, they occur at energies where
there are sharp features in the effective area curve due to instrumental
properties.  Improved calibration may therefore change our fits, but
we have been conservative in our interpretation of the
X-ray data, and we estimate that these changes will
only make the conclusions we have
drawn stronger.  We fit the spectrum with both Morrison \& 
McCammon (1983)
and Wilms, Allen \& McCray (2000) cross-sections, 
and obtain similar residuals for both.

If we fix the redshift of the absorber to be $z=0.312$ for
PKS 1127-145, then the absorber has N(HI)=1.2 $\pm$ 0.08 $\times$ 10$^{21}$ 
cm$^{-2}$ if the abundance is solar, compared to N(HI) = 5.1 $\pm$ 0.9 
cm$^{-2}$ derived from the damped Ly$\alpha$ profile (Lane et al. 1998).
The X-ray absorption implies that the abundance is about 23$\%$ solar, 
consistent with that found from the nebular emission lines.  

With the low spectral resolution of the ACIS-S, the redshift of the X-ray
absorption is not certain to be that of the intervening 21-cm absorption.
The two parameters of the absorber, redshift and N$_{H}$ are correlated: 
higher redshift requires higher N$_{H}$ to fit the observed spectrum.
Figure 4 shows the confidence levels for the two parameters, where the
1$\sigma$ contour has an elongated shape.  
When we allow the redshift of the absorber to vary, the best fit redshift
is $z\sim0.3$, but the chi-squared minimum is very shallow, and adequate
fits are allowed at other redshifts.  Better calibration of ACIS below 0.4 keV 
is needed to constrain the redshift better.  A grating observation would give
a definite answer about the location of the absorber, because of the possible
detection of the oxygen edge in the high resolution data.

However, the most plausible redshift for the X-ray absorber is $z=0.312$.
The UV spectrum shows that the $z=0.312$ absorber has by far the
highest column of H~I of any redshift system along the line of sight (Figure 5).
There is an absorber at $z_{abs}\sim z_{em}$, but it is weak, and has no
associated O VI or N V absorption;  it is likely a low column density Ly$\alpha$
forest cloud.  

We can rule out with high confidence an origin for the absorption in 
diffuse Milky Way interstellar gas.  We refit the X-ray data using
a power law and $z=0$ absorption only, 
with the column density allowed to vary as a free parameter.  The
best fit reduced $\chi ^2$ = 1.31 (517 degrees of freedom), compared
to $\chi ^2$ = 1.125 for the case with a $z_{abs} =$ 0.312 absorber.
Further, the best fit column if all the absorption is at $z=0$ is
1.35 $\pm$ 0.07 $\times 10^{21}$ cm$^{-2}$, compared to 0.383 $\pm$ 0.01 
$\times 10^{21}$ cm$^{-2}$ measured from high resolution 21-cm emission
data by Murphy et al. (1996).  The absorption column required 
to fit the X-ray data is more than 10$\sigma$ greater than the Galactic
column, and is unlikely to be at $z=0$.

\section{Limits on Abundances of Individual Elements for the PKS1127-145 Absorber}

The spectral resolution of the ACIS-S is insufficient to resolve
the individual edges of the elements. 
Moreover, we expect that most of the opacity is from oxygen, 
and if the absorption is at $z=0.321$, 
the K-edge of oxygen at
rest energy 0.532 keV is redshifted to 0.405 keV, 
outside the part of the ACIS-S pass-band with reliable calibration.
We tried a fit with $z_{abs}$=0.312 frozen, and two absorbers, one
with He/H only, frozen at the primordial value, and the other with 
$``$processed" metals, including oxygen (c.f. Madesjki et al. 1996). We 
allowed the total column of each to vary.
The results are shown figure 5.  

We found that we could fit the data with a model where 
all the absorption is 
the result of helium (and N$_{H}$=$3.31\pm 0.15 \times 10^{21}$ cm$^{-2}$),
about a factor of 3 greater than the column derived assuming heavy 
metals are present. 
It is implausible however 
that the X-ray absorber would be pure helium and hydrogen,
since the $z_{abs}=$ 0.312 system shows strong metal absorption in the 
ultraviolet. 

Even if the X-ray absorption opacity is due to helium and hydrogen only,
we can still place a limit on the oxygen abundance relative to 
the H I measured by the damped Ly$\alpha$ profile.  The 3$\sigma$ upper 
limit to the oxygen abundance
is 17$\%$ solar.  Improved calibration of ACIS-S may change
these results significantly.  Grating observations with Chandra will be able
to measure the oxygen edge directly, and remove the ambiguity 
in the assumed ratio of the elements heavier than hydrogen.

\section{Discussion}

High quality spectra from ground and from HST have been used
to investigate the detailed abundance patterns in damped Ly$\alpha$
absorbers (Lauroesch, Truran, Welty \& York 1996; 
Pettini et al. 1999 and references therein).  
The observed pattern
appears to be consistent with
the Galactic halo  pattern (Population II) with little
dust depletion, although dust depleted solar ratios cannot be
ruled out (Lu et al. 1996; Prochaska \& Wolfe 1999), 
although uncertainties 
remain. 
Saturation of the absorption lines, particularly for the more abundant
elements like C, N and O mean that 
column densities of these elements 
are difficult to determine (Lu, Sargent \& Barlow 1998).
Elements condense out of the gas phase onto dust grains, so that
the absorption line columns are
estimated to be low by factors of  3-5 for lightly refractory
elements such as C, N and O, and depleted by factors of $\sim$1000 or so for
Fe, Mg or Si (Jenkins, Savage \& Spitzer 1986).  The exact ratio depends
on details of the grain chemistry, and the particular history of shocks
which the material has suffered.
Wide variations in the depletion factors are seen empirically in
Milky Way ISM clouds (see Tielens 1998, and references therein).

One element which is not depleted onto dust, however, is zinc 
(Morton 1974; de Boer, Fitzpatrick, \& Savage 1985; 
Sembach, Steidel, Macke, \& Meyer 1995).
and 
extensive surveys of the abundance of Zn II relative to hydrogen in
damped Ly$\alpha$ absorbers (Meyer, Lanzetta, \& Wolfe 1995,
Pettini et al. 1997).
Zinc shares the same nucleosynthetic
origin as iron and so Zn/H gives a reliable indication of Fe/H.
Iron unfortunately suffers heavy, variable depletion onto dust
grains and therefore is impossible to use to measure abundances 
(deBoer \& Lamers 1978).

The signature of Population II abundances is an enhancement of
oxygen and other even-proton nuclei (Ne, Mg, Si, S, Ar and Ca)
relative to the iron group elements (Fe, Zn, Cr) (Lauroesch 
et al. 1996; Timmes, Lauroesch \& Truran 1995 and
references therein).  Such
an enhancement is observed
very clearly in the analysis of the oldest, metal-poor halo stars in the Milky
Way where the oxygen group (by which we mean O, Ne, Mg, Si, S, Ar and Ca)
is enhanced by a factor of 3 relative to solar ratios of iron and other
even-Z iron peak nuclei (Zn, Fe, Cr, Ni).   The reason is that
oxygen, along with other $\alpha$-elements, is produced
primarily by massive stars which evolve on short time scales
compared with the lower-mass progenitors of Type Ia SNe which
produce the iron group elements such as zinc.  It is therefore
expected that O/H will increase towards the solar value
earlier than Zn/H, particularly given the high star-formation rates
expected in the early stages of galaxy evolution.

The X-ray absorption detected with Chandra
for PKS 1127-145 
measures the abundance of the oxygen group
elements directly.
The X-ray absorbing column measures the metallicity of the
gas independent of dust depletion and saturation effects, and
is caused by exactly those elements (O, Ne, Mg, Si, S, Ar. Ca) which
are expected to be enhanced in Pop II objects (see Morrison and McCammon
(1983).  By comparing metallicity
derived from the X-ray absorbing column with that measured by the
Zn II lines (iron group) we can see directly if the damped Ly$\alpha$
has a halo-type abundance pattern, or a disk-type abundance pattern.
These observations also serve as an important check on the
UV absorption line spectroscopy.

The existing HST spectra of PKS 1127-145 unfortunately 
are inadequate to detect
Zn II at the strength predicted by our X-ray measurements. 
Ultraviolet transitions of S II are also interesting to search for, 
but fall within the Ly$\alpha$ forest;  
the strongest S II transition at $\lambda$1259 is also  blended
with the much stronger Si II $\lambda$1260 line.  Future high resolution
observtions in the UV may be able to detect S II lines for direct comparison 
to the oxygen abundance derived from the X-ray data. 
Spectroscopy to search for Zn II absorption with the echelle on STIS is planned.

\acknowledgements
\leftline{ACKNOWLEDGMENTS}

We thank the staff of the Chandra X-ray Center  
for their aid with the planning and analysis 
of these observations. We thank Greg Madejski for suggesting we  
consider the relative importance of absorption by helium.
JB appreciates financial support
from NSF grant AST-9617060 and Chandra Guest Observer grant 
GO0-1164X from NASA.  AS, ME, TA and AD acknowledge support 
from NASA contract No.\ NAS8-39073 (ASC).  
This research has made use of the NASA/IPAC Extragalactic Database 
(NED) which is operated by the Jet Propulsion Laboratory, 
California Institute of Technology, under contract with NASA, as well as 
 data obtained with the the NASA/ESA Hubble Space Telescope, 
retrieved from the archive at the Space Telescope Science Institute, 
which is operated by the Association of Universities for Research in 
Astronomy, Inc. under NASA contract No. NAS5-26555.

\newpage


\begin{figure}
\figurenum{1}
\plottwo{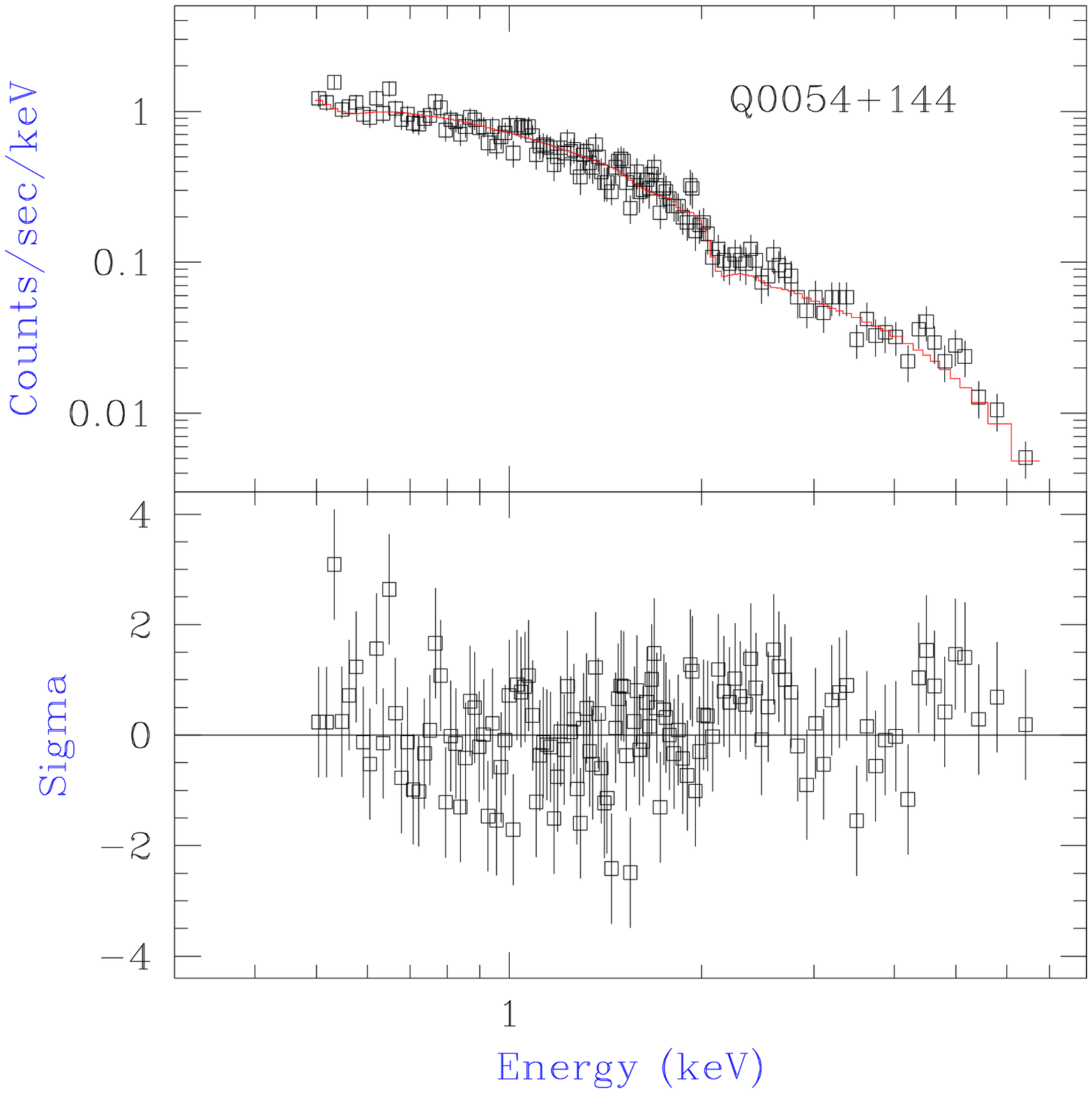}{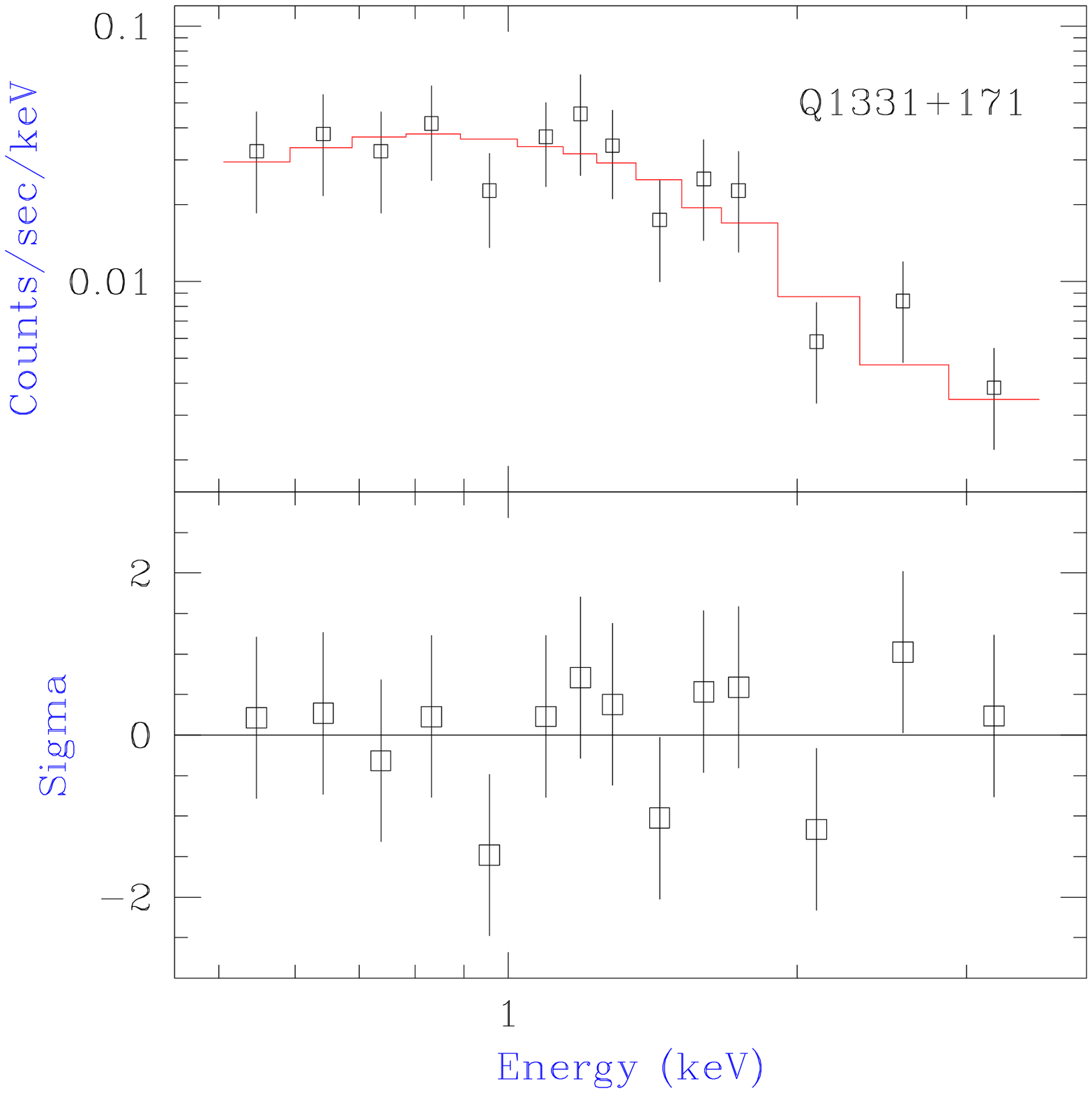}
\caption{$Chandra$ ACIS-S spectra of Q 0054+144 and Q 1331+171.
Bottom panel shows residuals to best fit power law, plus
Galactic absorption.  No intervening absorption is necessary
to fit the spectra, although the limits on the column 
densities for the intervening
absorbers are weak. 
}
\end{figure}

\begin{figure}
\figurenum{2}
\plotone{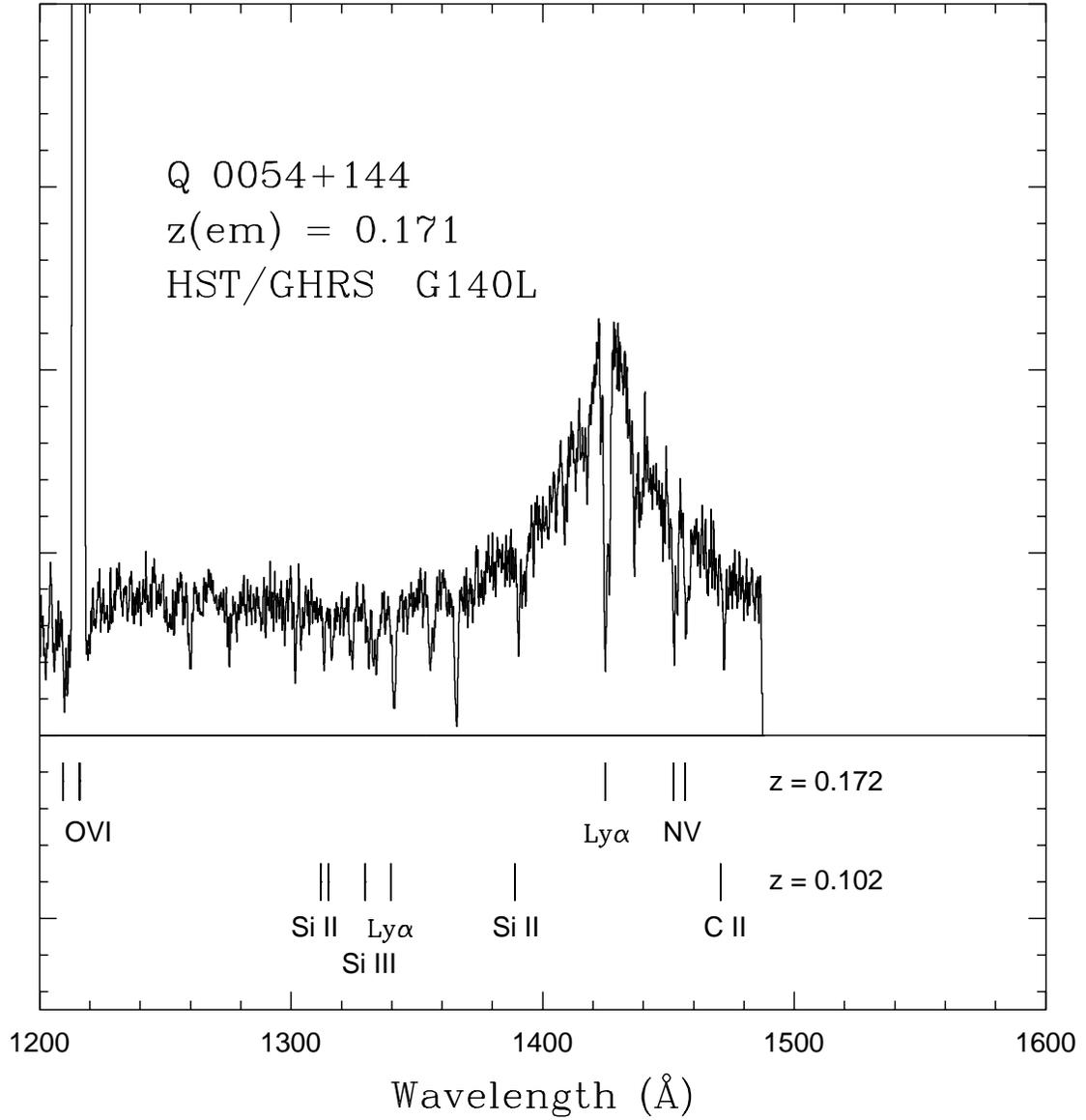}
\caption{$HST$ GHRS spectrum of Q0054+144, obtained 
with the G140L grating. The expected transitions of
the $z_{abs}=0.172$ associated absorber and $z_{abs}$= 
0.102 absorber are marked.  The broad emission feature
is Ly$\alpha$.
}
\end{figure}

\begin{figure}
\figurenum{3}
\plottwo{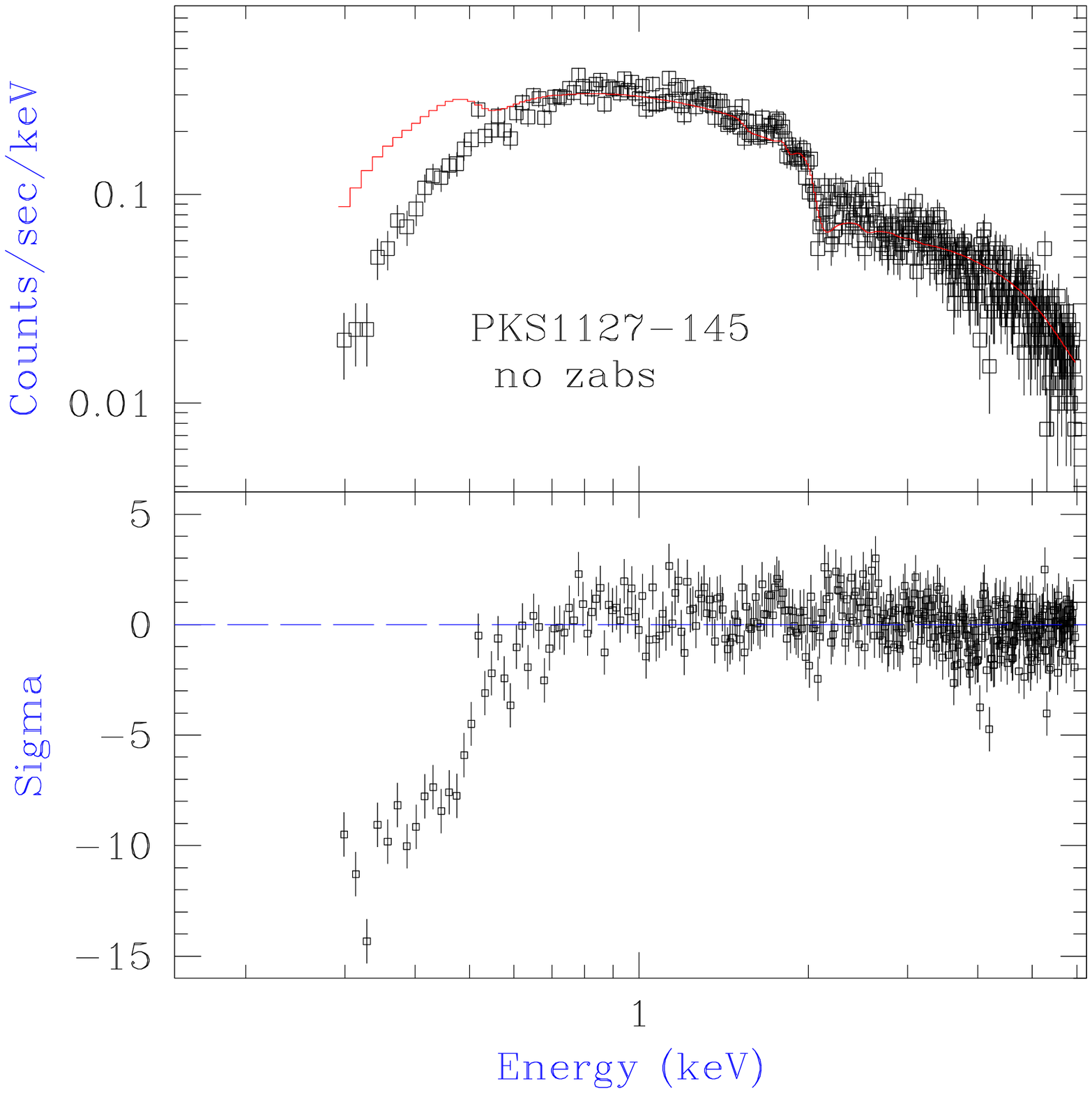}{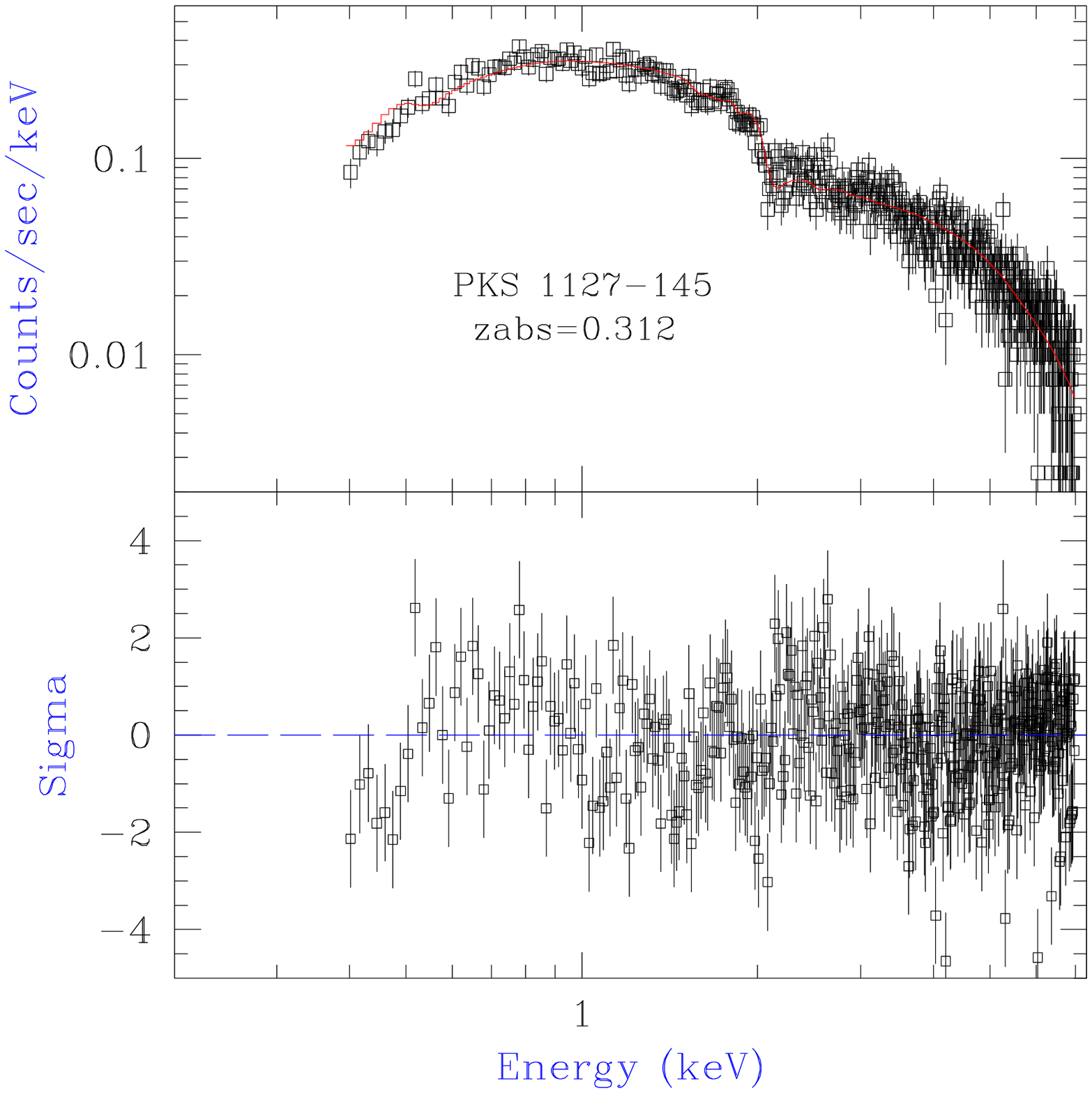}
\caption{$Chandra$ ACIS-S spectrum of $z_{em}$=1.17 quasar 
PKS 1127-145.  Left panel
shows best fit to power-law and Galactic absorption only. 
Right
panel shows best fit to power-law, Galactic absorption, and 
intervening absorption with fixed redshift $z=0.312$.} 
\end{figure}

\begin{figure}
\figurenum{4}
\plotone{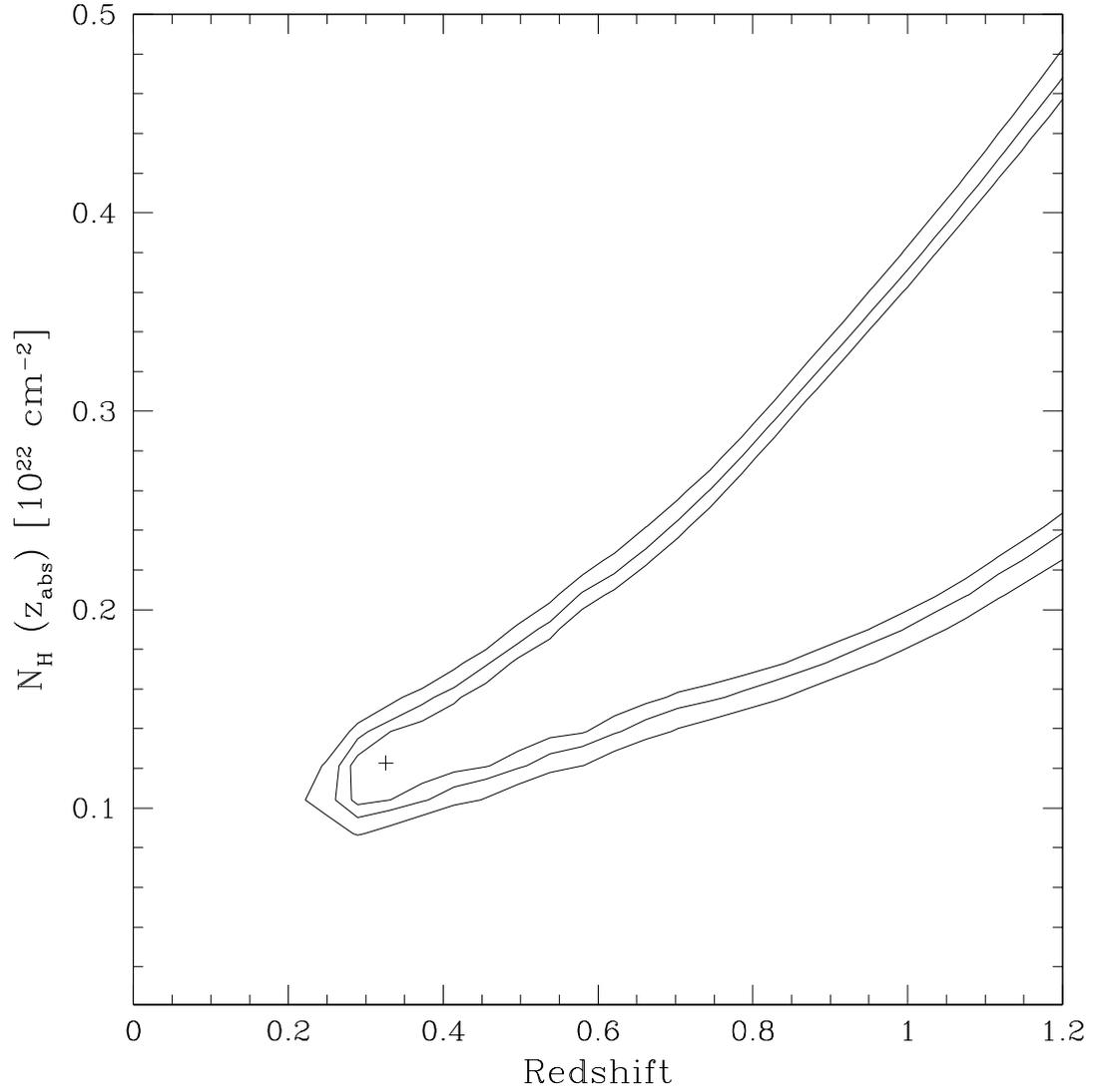}
\caption{99.7$\%$, 95.5$\%$, 68.3$\%$ confidence levels
for two parameters (N$_{H}$ versus $z_{abs}$) of the power law and
redshifted absorber model fit to PKS 1127-145 data. The best fit
parameters are indicated with the cross.}
\end{figure}

\begin{figure}
\figurenum{5}
\plotone{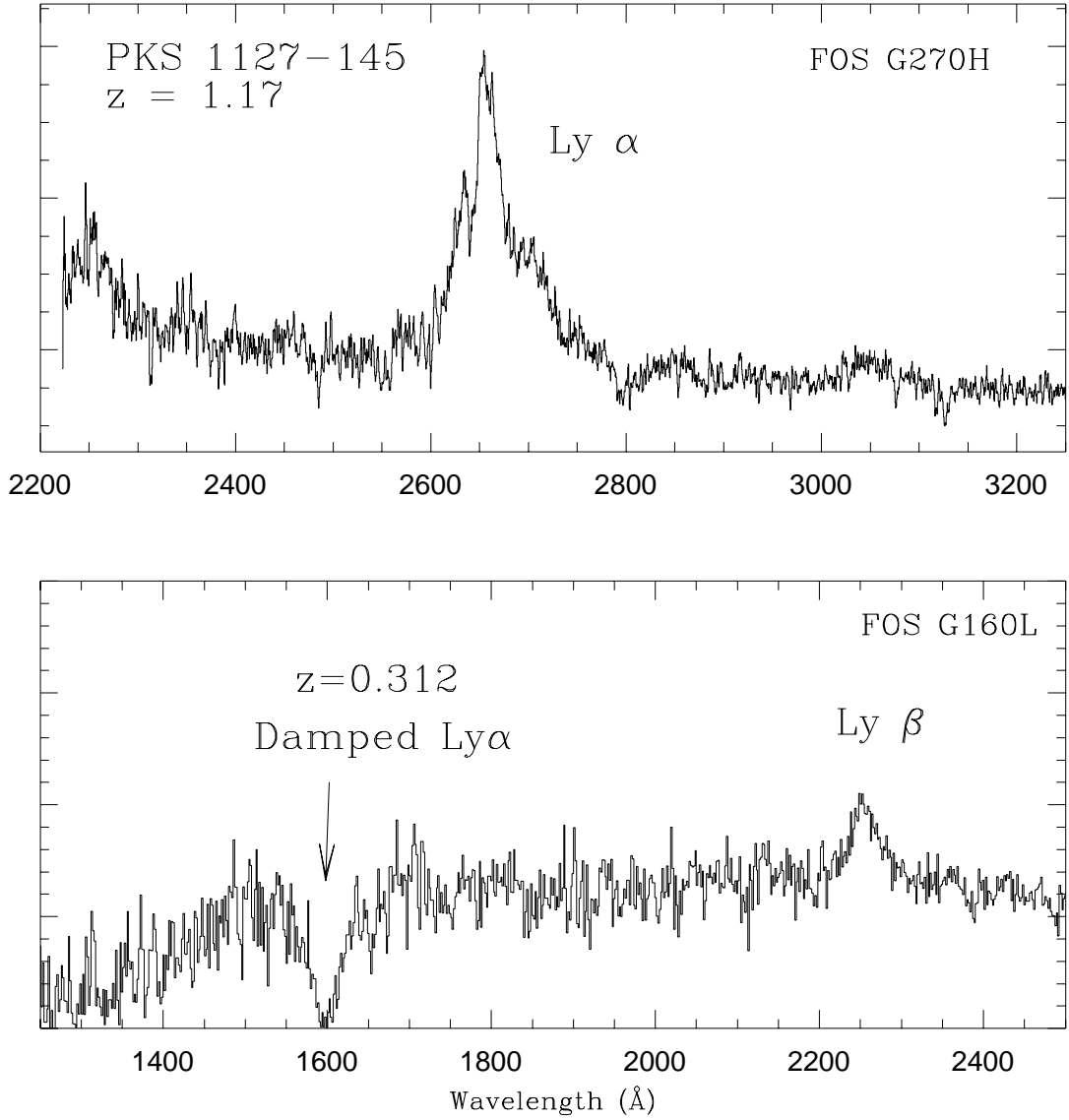}
\caption{FOS Spectra of PKS 1127-145, $z_{em}$=1.17. The prominent 
damped Ly$\alpha$ absorber at $z_{abs}$=0.312 has the highest column
density of neutral hydrogen of redshift system in the spectrum. 
}
\end{figure}

\begin{figure}
\figurenum{6}
\plotone{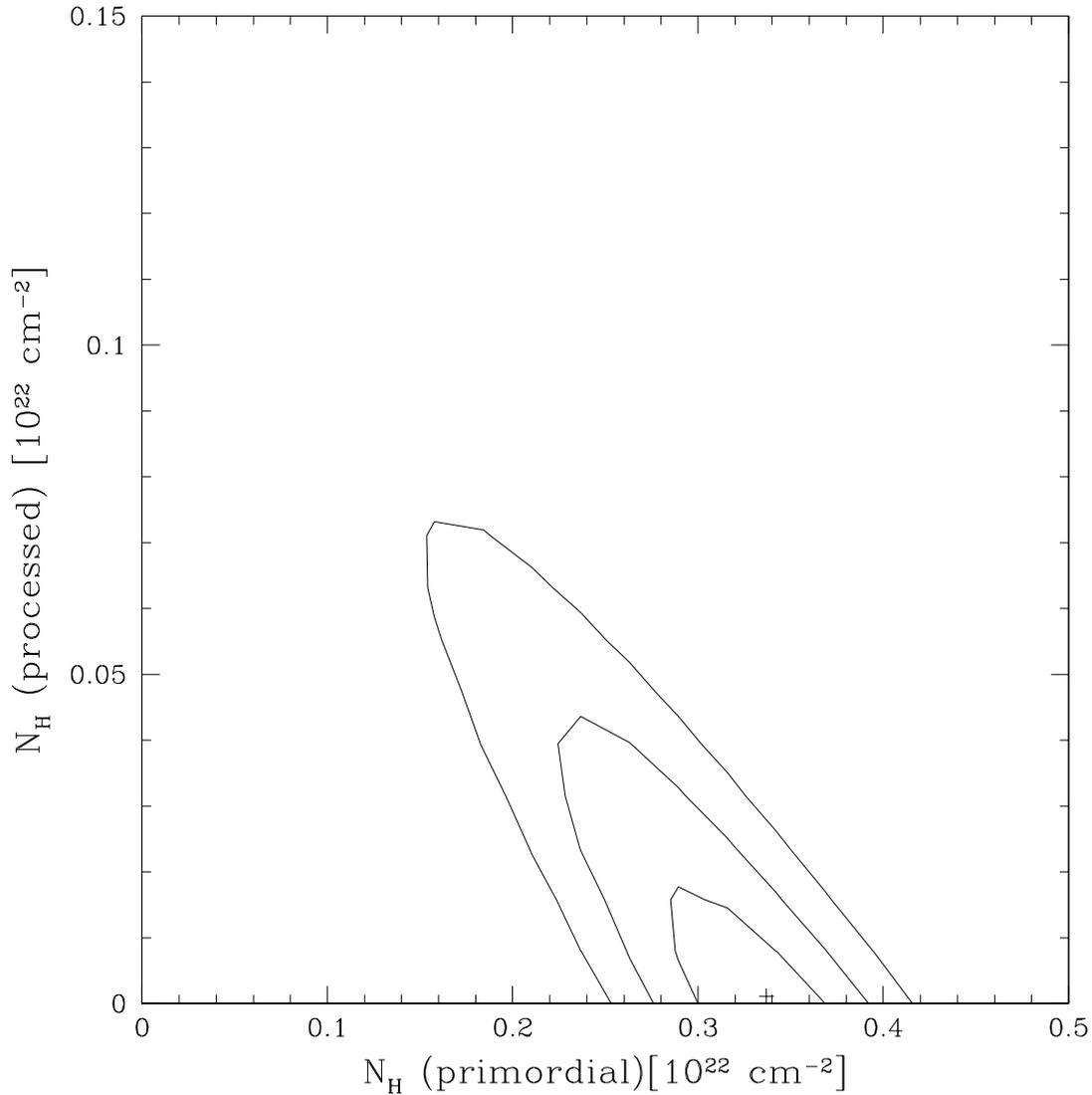}
\caption{99.7$\%$, 95.5$\%$, 68.3$\%$ confidence levels
for two parameters $``$primordial" N$_{H}$ (i.e. pure helium) 
versus $``$processed" N$_{H}$ (mostly oxygen), for the ACIS-S
data of PKS 1127-145.  The fits assumed a power law, fixed Galactic
absorption, and a redshifted absorber with $z_{abs}$=0.312. 
The best fit
parameters are indicated with the cross, and imply an upper limit
to the O/HI of 17$\%$.}
\end{figure}

\clearpage

\begin{deluxetable}{lccrclr}
\doublespace
\tablecaption{Observations}
\tablehead{
\colhead{Object} & \colhead{$z_{em}$} & \colhead{Date} &  
\colhead{Exp.$^{a}$} & \colhead{N$_{HI}$$^{b}$} & 
\colhead{$\alpha_{x}$, $\delta_{x}^{c}$} & \colhead{Counts$^{d}$}  \\
\colhead{} & \colhead{} & \colhead{Seq. No., OBSID} & 
\colhead{($s$)} & \colhead{} & 
\colhead{(J2000)} & \colhead{}  \\
}
\startdata
Q 0054+144   & 0.171  & 2000 July 29 & 4660 & 4.35 & 00:57:09.99     & 6210 \\
	     &        & 700170, 865  &      &       & +14:46:10.46 & \\
PKS 1127-145 & 1.187  & 2000 May 28 & 27,358 & 3.83 & 11:30:07.014 & 16,238 \\
             &        & 700171, 866  &        &      & -14:48:27.35 & \\ 
Q 1331+170   & 2.08   & 2000 April 3 & 3021 & 1.8 & 13:33:35.708 & 198 \\
             &        & 700172, 867  &        &      & +16:49:03.73 & \\
\enddata
\tablenotetext{a} {Exposure time in seconds, with 0.9068 dead time correction
applied.}
\tablenotetext{b} {Galactic neutral hydrogen column density, 
10$^{20}$ cm$^{-2}$, from Stark et al. (1992) for Q 0054+144 
and Q 1331+170, and Murphy et al. (1996) for PKS 1127-145.}
\tablenotetext{c}{Right ascension, declination of X-ray source.}
\tablenotetext{d}{Counts in circle of radius 2.5 \arcsec, 0.4 - 8.0 keV.  
}
\end{deluxetable}

\begin{deluxetable}{lccccr}
\doublespace
\tablecaption{Results of Spectral Fitting}
\tablehead{
\colhead{Object} & \colhead{$z_{abs}^{a}$} & \colhead{$N_H ^{b}$} &
\colhead{$\Gamma^{c}$} & \colhead{Norm$^{d}$} & \colhead{$f_x^{e}$}    \\
}
\startdata
Q 0054+144   & 0.171  & $<$2.8        & 1.95$\pm$0.04  & 13.29$\pm$0.04& 36.7 
\\
             & None   &   -- & 1.97$\pm$0.03  & 13.24$\pm$0.18& 36.8 
\\
PKS 1127-145 & 0.312  & 1.19$\pm$0.08 & 1.19$\pm$0.02  & 6.48$\pm$0.15 & 59.7 
\\
             & 1.187  & 3.41$\pm$0.21 & 1.17$\pm$0.02  & 6.28$\pm$0.11 & 60.3 
\\
             & None   &   -- & 0.98$\pm$0.01  & 5.10$\pm$0.08 & 68.0 
\\
	     & Free &1.198$\pm$0.061& 1.21$\pm$0.02  & 6.67$\pm$0.06 & 59.7 
\\
Q 1331+170   & 1.77   & $<$19.1       & 1.48$\pm$0.19  & 0.54$\pm$0.09 & 3.14 
\\
             & None   &  -- & 1.48$\pm$0.13  & 0.54$\pm$0.05 & 3.13 
\\
\enddata
\tablenotetext{a} {Fits assuming a power law, in all cases Galactic 
absorption was fixed. 
$``$None" means no additional redshifted absorption was included.  
$``$Free" means that the fits included 
redshifted absorption, with $z_{abs}$ allowed to 
vary.  Other fits included redshifted absorption  
at fixed redshift z$_{abs}$, with N$_H$ of the redshifted absorber 
allowed to vary.}
\tablenotetext{b} {Redshifted absorption column, 10$^{21}$ cm$^{-2}$.  
One-sigma errors for PKS1127-145; 3-sigma upper limit for Q1331+170 and Q 
0054+144.
}
\tablenotetext{c} {Photon spectral index, with one-sigma error.}
\tablenotetext{d} {Power law normalization and one-sigma error,
10$^{-4}$photons cm$^{-2}$ s$^{-1}$ keV$^{-1}$, at 1 keV.}
\tablenotetext{e} {unabsorbed flux, 2-10 keV, 10$^{-13}$ ergs cm$^{-2}$ 
s$^{-1}$}
\end{deluxetable}

\end{document}